\documentclass[aps,prd,nofootinbib,twocolumn,groupaddress,preprintnumbers]{revtex4-2}

\usepackage{amssymb}
\usepackage{amsmath}
\usepackage{epsfig}
\usepackage{breakurl}
\usepackage{bm}
\usepackage{color}
\usepackage{tikz}

\makeatletter

\usepackage[utf8]{inputenc}
\usepackage{dsfont}
\usepackage{shuffle}
\usepackage{array}
\usepackage{amssymb,amsmath}
\usepackage[margin=2.2cm]{geometry}
\usepackage{mathtools}
\usepackage{bbold}
\usepackage{color}

\usepackage[
      colorlinks=true,
      linkcolor=darkblue,  
      urlcolor=blue,    
      filecolor=blue,     
      citecolor=red,
      linktocpage=true,
      pdfstartview=FitV,
      bookmarksopen=true    
      ]{hyperref}

\definecolor{darkblue}{rgb}{0.2, 0, 0.8}
\definecolor{darkred}{rgb}{0.9, 0, 0.1}
\usepackage{tikz}
\usetikzlibrary{arrows,shapes}
\usetikzlibrary{trees}
\usetikzlibrary{patterns}
\usetikzlibrary{matrix,arrows} 				% For commutative diagram
\usetikzlibrary{positioning}				% For "above of=" commands
\usetikzlibrary{calc,through}				% For coordinates
\usetikzlibrary{decorations.pathreplacing}  % For curly braces
\usepackage{pgffor}							% For repeating patterns

\usetikzlibrary{decorations.pathmorphing}	% For Feynman Diagrams
\usetikzlibrary{decorations.markings}
\tikzset{
    vector/.style={decorate, decoration={snake}, draw},
	provector/.style={decorate, decoration={snake,amplitude=2.5pt}, draw},
	antivector/.style={decorate, decoration={snake,amplitude=-2.5pt}, draw},
        smallvector/.style={decorate, decoration={snake,amplitude=1.5pt,post length=0.5mm}, draw},
    fermion/.style={draw=black, postaction={decorate},
        decoration={markings,mark=at position .55 with {\arrow[draw=black]{>}}}},
    fermionbar/.style={draw=black, postaction={decorate},
        decoration={markings,mark=at position .55 with {\arrow[draw=black]{<}}}},
    fermionnoarrow/.style={draw=black},
    gluon/.style={decorate, draw=black,
        decoration={coil,amplitude=4pt, segment length=5pt}},
    scalar/.style={dashed,draw=black, postaction={decorate},
        decoration={markings,mark=at position .55 with {\arrow[draw=black]{>}}}},
    scalarbar/.style={dashed,draw=black, postaction={decorate},
        decoration={markings,mark=at position .55 with {\arrow[draw=black]{<}}}},
    scalarnoarrow/.style={dashed,draw=black},
    electron/.style={draw=black, postaction={decorate},
        decoration={markings,mark=at position .55 with {\arrow[draw=black]{>}}}},
    bigvector/.style={decorate, decoration={snake,amplitude=4pt}, draw},
    arrow/.style={draw=black, postaction={decorate},
        decoration={markings,mark=at position 1 with {\arrow[draw=black]{>}}}},
}

\tikzstyle{block} = [draw, rectangle, 
    minimum height=3em, minimum width=6earticlem]

\newcommand{\reef}[1]{(\ref{#1})}

\def\be{\begin{equation}}
\def\ee{\end{equation}}
\def\bea{\begin{eqnarray}}
\def\eea{\end{eqnarray}}
\def\ba{\begin{array}}
\def\ea{\end{array}}
\def\bd{\begin{displaymath}}
\def\ed{\end{displaymath}}

\def\>{\rangle} %right angle
\def\<{\langle} %left angle
\def\Dsl{D \hskip-.6em \raise1pt\hbox{$ / $ } }
\def\to{\rightarrow}

\newcommand{\bes}{\begin{split}}
\newcommand{\ens}{\end{split}}

\begin{document}
\title{Bootstrapping the String KLT Kernel}
\author{Alan Shih-Kuan Chen}
\author{Henriette Elvang}
\author{Aidan Herderschee}
\affiliation{Leinweber Center for Theoretical Physics, Randall Laboratory of Physics\\  University of Michigan, 450 Church St, Ann Arbor, MI 48109-1040, USA}
\email{shihkuan@umich.edu,\,elvang@umich.edu,\,aidanh@umich.edu}

\begin{abstract}

We show that a generalized version of the 4-point string theory KLT double-copy map is the most general solution to the minimal-rank double-copy bootstrap in effective field theory. This follows from significant restrictions of the 4-point map resulting from the 6-point bootstrap analysis. The generalized 4-point double-copy map is defined by a function with only two parameters times a simple function that is symmetric in $s,t,u$. The two parameters can be interpreted as independent choices of $\alpha'$, one for each of the two sets of amplitudes double-copied with the map. Specifically, each of those two sets of amplitudes must obey either the string monodromy relations or the field theory KK \& BCJ relations; there are no other options. We propose a closed form of the new double-copy map that interpolates between the original KLT string double-copy and the open \& closed string period integrals. The construction clarifies the ``single-valued projection" property of the Riemann zeta-function values for the 4-point string theory double copy.

\end{abstract}

\preprint{LCTP-23-01}

\maketitle

\section{Introduction} 

The double copy is a map between observables in a variety of theories, including, famously, Yang-Mills theory (YM) to gravity. It originates in string theory as a relation between open and closed string amplitudes \cite{Kawai:1985xq} and has, in its field theory incarnation, dramatically simplified higher-order amplitude computations. 
The double copy has a wide range of applications, such as derivations of compact formulas for tree-level amplitudes at $n$-point \cite{Bern:2008qj,Bern:2010ue,Bern:2019prr,Elvang:2007sg,Cachazo:2013iea,Dolan:2013isa,Cachazo:2013gna,Cachazo:2013iaa,Cachazo:2014nsa,Cachazo:2014xea,Mizera:2017rqa} and the simplification of higher-loop calculations to examine, for example, the UV behavior of supersymmetric scattering amplitudes \cite{Bern:1998ug, Bern:1998xc, Bern:1998sv,Bern:2006kd,
Bern:2007xj,Bern:2008pv,Bern:2009kd,Bern:2012gh,Bern:2012uf,Bern:2013uka,Bern:2014sna,Bern:2015ooa,Bern:2017ucb,Bern:2018jmv}. It is used in investigations of gravitational wave physics \cite{Bern:2019nnu,Bern:2019crd}, studies of the interplay between quantum field theory and string theory \cite{Mason:2013sva,Berkovits:2013xba,Geyer:2014fka,Mizera:2017cqs,Arkani-Hamed:2017mur,Mizera:2019gea,Mizera:2019blq}, methods for finding exact solutions to classical equations of motion \cite{Monteiro:2014cda,Luna:2015paa,White:2016jzc,Luna:2016due,Goldberger:2016iau,Luna:2016hge,Goldberger:2017frp,Arkani-Hamed:2019ymq,Cheung:2022mix}, computations of boundary observables in (anti)-de Sitter \cite{Li:2018wkt,Fazio:2019iit,Armstrong:2020woi,Albayrak:2020fyp,Zhou:2021gnu,Diwakar:2021juk,Cheung:2021zvb,Cheung:2022pdk,Herderschee:2022ntr,Drummond:2022dxd,Li:2022tby,Bissi:2022wuh,Lee:2022fgr}, explorations of amplitudes in the soft and Regge limit \cite{Saotome:2012vy,SabioVera:2012zky,SabioVera:2014mkb,Oxburgh:2012zr}, and studies of how different symmetries manifest at the level of amplitudes \cite{Elvang:2010kc,Bern:2017rjw,Bern:2017tuc,Bern:2017rjw,Johansson:2018ues,Elvang:2018dco,Ben-Shahar:2018uie,Elvang:2020kuj,Carrasco:2022jxn,Monteiro:2022nqt}. Reviews of the double-copy include Refs.~\cite{Bern:2002kj,Bern:2019prr,Bern:2022wqg,Adamo:2022dcm}.

In this Letter, we study the double copy in the context of $d$-dimensional Effective Field Theory (EFT) with massless states and local higher-derivative operators. 
Ref.~\cite{Chi:2021mio} proposed an algorithm, the {\em KLT double-copy bootstrap}, to systematically calculate constraints on higher-derivative corrections to the field theory double-copy map. 
When implemented at minimal rank  (motivated by absence of spurious poles) 
for 4- and 5-point amplitudes, Ref.~\cite{Chi:2021mio} found that the bootstrap gives an EFT double-copy map that is more general than, but includes the string kernel. 

This Letter extends the minimal rank KLT bootstrap to 6-point, finding novel restrictions on the 4- and 5-point double-copy maps that leave very few parameters free. As a consequence, up to and including 10-derivative order, we find that {\em any EFT amplitude compatible with the 4-point double-copy map must obey either the string monodromy relations or the field theory KK and BCJ relations} \cite{Bern:2008qj,Plahte:1970wy,Bjerrum-Bohr:2009ulz,Stieberger:2009hq,Bjerrum-Bohr:2010mia,Bjerrum-Bohr:2010pnr}. It is surprising that string monodromy arises in an EFT context because it is a property of the worldsheet description of string scattering \cite{Plahte:1970wy,Bjerrum-Bohr:2009ulz,Stieberger:2009hq,Bjerrum-Bohr:2010mia,Bjerrum-Bohr:2010pnr}. 
We extend the bootstrap results to 36-derivative order using the hybrid decomposition conjecture in Ref.~\cite{ACHElater} and the results of Ref.~\cite{Chen:2022shl}.

Based on the low-energy expansion of the 
generalized double-copy map, we propose a closed-form expression for the 4-point map: it takes a factorized form of one function that depends on two parameters (a ``left" and a ``right" choice of $\alpha'$) and a symmetric function $U$ in $s,t,u$ of a particularly simple form.  
Up to this symmetric function, the most general double-copy map is a generalization of the string theory double copy such that a continuous interpolation is possible between the string kernel with $\alpha'_\text{L} = \alpha'_\text{R} \ne 0$,  the open string period integrals (``Z-theory") 
with either $\alpha'_\text{L}$ or $\alpha'_\text{R}$ zero, and the closed string period integrals with $\alpha'_\text{L} = \alpha'_\text{R} = 0$. 
In an EFT context, the double copy is effectively independent of the symmetric function $U$ in the sense that $U$ can be absorbed into the input amplitudes. 
In a string theory context, the symmetric function provides insight into the single-valued projection property of closed string amplitudes \cite{Schnetz:2013hqa,Brown:2013gia}.

%%%%%%%%%%%%%%%%%%%%%%%%%%
%%%%%%%%%%%%%%%%%%%%%%%%%%
%%%%%%%%%%%%%%%%%%%%%%%%%%
\section{Review: KLT Bootstrap}\label{review}
%%%%%%%%%%%%%%%%%%%%%%%%%%
%%%%%%%%%%%%%%%%%%%%%%%%%%
%%%%%%%%%%%%%%%%%%%%%%%%%%

The KLT double-copy maps on-shell tree amplitudes, $A_n^\text{L}$ and $A_n^\text{R}$, of two theories L and R to on-shell tree amplitudes in a third theory, denoted $\text{L} \otimes \text{R}$. Assuming  all states are massless, we have
\begin{equation}
\label{KLT}
    {A}_n^{\text{L}\otimes \text{R}} = \sum_{a \in {B}_\text{L},\,b \in {B}_\text{R}} 
    {A}_n^{\textrm{L}}[a]\,
    S_n[a|b]\,
    {A}_n^{\textrm{R}}[b]\,.
\end{equation}
 The sum is over two ``basis choices" ${B}_\text{L}$ and $B_\text{R}$ of $(n-3)!$ color-orderings among the $(n-1)!$ single-trace structures associated with a local or global color-group of the L and R theories.  
The {\em double-copy kernel} $S_n$ is a function of the $n$-point Mandelstam variables. In the original KLT construction of Ref.~\cite{Kawai:1985xq}, the L and R amplitudes are open string tree amplitudes and $S_n$ is the {\em string kernel}. In the low-energy limit $\alpha' \to 0$, the open (closed) string amplitudes reduce to YM (gravity) tree amplitudes and the string kernel becomes  the {\em field theory kernel}. 

A key property of Eq.~\reef{KLT} is that the double-copied amplitudes, ${A}_n^{\text{L}\otimes \text{R}}$, are independent of the choice of color-orderings, ${B}_\text{L/R}$, in the sum. As a consequence, the double copy can  be formulated as a multiplicative map of field theories,  
\be
\label{FTmap}
 \text{FT}^{\text{L}\otimes \text{R}}  = \text{FT}^\text{L}  \otimes  \text{FT}^\text{R} \,,
\ee
where the multiplication rule $\otimes$ is defined by a kernel $S_n$. In string theory, basis independence is ensured by the string monodromy relations satisfied by the open string tree amplitudes. In field theory, the KK and BCJ relations, denoted as {\em KKBCJ relations}, guarantee the needed basis independence.

The goal of the {\em KLT double-copy bootstrap} introduced in Ref.~\cite{Chi:2021mio} is to determine the most general form of the double-copy kernel, $S_n$, along with the associated linear relations required by the L and R amplitudes such that the double-copy is a map on the space of field theories. This requires that 
the output ${A}_n^{\text{L}\otimes \text{R}}$ is free of spurious poles and independent of the basis choices ${B}_\text{L/R}$ in Eq.~\reef{KLT}. Ref.~\cite{Chi:2021mio} showed that a fundamental principle of the double-copy map is the existence of an identity element $\mathbb{1}$, i.e.~a field theory 
whose tree amplitudes obey 
\bea
  \label{KLTalgebra0}
     &&\mathbb{1} \otimes \mathbb{1} = \mathbb{1}\,, \hspace{5mm} \\[2mm]
  \label{KLTalgebra0cond}
     &&\text{L}\otimes \mathbb{1} = \text{L}\,, \hspace{5mm} 
     \mathbb{1} \otimes  \text{R} =  \text{R}   
     \,.
\eea
This is the {\em KLT algebra}. The identity model associated with the field theory kernel is the cubic bi-adjoint scalar model (BAS),
 \be
 \label{BAS}
   \mathcal{L}_\text{BAS} = -\frac{1}{2} (\partial \phi^{aa'})^2 + \frac{1}{6} f^{abc} \tilde{f}^{a'b'c'} \phi^{aa'} \phi^{bb'} \phi^{cc'} \,,
\ee
for which Eq.~\reef{KLTalgebra0cond} becomes the KKBCJ relations. 

The string theory kernel has an associated identity model whose `amplitudes' were constructed in Ref.~\cite{Mizera:2016jhj}. In the low-energy $\alpha'$-expansion, they are tree amplitudes of the BAS model plus a particular tower of higher-derivative terms with fixed coefficients controlled by $\alpha'$. 

The core proposal of Ref.~\cite{Chi:2021mio} is to take the KLT algebra to be a fundamental property of the double-copy. Specifically, $\mathbb{1} \otimes \mathbb{1} = \mathbb{1}$ provides a way to bootstrap the kernel in terms of the identity model and generalized KKBCJ relations arise from Eq.~\reef{KLTalgebra0cond}.   
As the candidate for the identity model, we consider a general BAS EFT, which includes the cubic interaction in Eq.~\reef{BAS} and a general ansatz for the higher-derivative terms. 
 
Let $m_n[a|b]$ be the doubly color-ordered tree amplitudes of the general  BAS EFT. The bootstrap equation $\mathbb{1} \otimes \mathbb{1} = \mathbb{1}$  says 
\begin{equation}
\label{KLT1}
    {m}_n[c|d] = 
    \sum_{a \in {B}_\text{L},\,b \in {B}_\text{R}} 
    {m}_n[c|a]\,
    S_n[a|b]\,
    {m}_n[b|d]\,.
\end{equation}
The sums are over the $(n-3)!$ color-orderings in the bases defined below Eq.~\reef{KLT}. Since there is a total of $(n-1)!$ color-orderings, we organize the amplitudes $m_n[a|b]$ and
the kernel components $S_n[a|b]$ 
into $(n-1)! \times(n-1)!$ matrices $\mathbf{m}_n$ and $\mathbf{S}_n$. If the color-choices $c$ and $d$ in Eq.~\reef{KLT1} run over $(n-3)!$ color-orderings $B_1$ and $B_2$, then Eq.~\reef{KLT1} becomes a matrix equation 
\be
  \label{mandS}
  \mathbf{m}_n^{B_1B_2} = \mathbf{m}_n^{B_1B_L}
  \,.\,
  \mathbf{S}_n^{B_LB_R}    \,.\,
  \mathbf{m}_n^{B_RB_2}\,,
\ee
where the superscript indicates the choices of rows and columns for the 
$(n-3)! \times (n-3)!$ sub-matrices of $\mathbf{m}_n$ and $\mathbf{S}_n$. Choosing $B_1=B_R$ and $B_2=B_L$, it follows from Eq.~\reef{mandS} that
\be  
\label{mandSinv}
\mathbf{S}_n^{B_L B_R} = \big(\mathbf{m}_n^{B_R B_L}\big)^{-1}\,.
\ee
This shows that {\em the kernel is uniquely linked to the tree amplitudes of the identity model}. Plugging Eq.~\reef{mandSinv} into Eq.~\reef{mandS} gives a non-trivial condition: the rank of the $(n-1)! \times(n-1)!$ matrix $\mathbf{m}_n$ must be $(n-3)!$. 
Generic BAS EFT tree amplitudes give $\mathbf{m}_n$ with full rank $(n-1)!$. Thus, the ``minimal rank" condition imposes nontrivial relations on the Wilson coefficients and $\mathbb{1} \otimes \mathbb{1} = \mathbb{1}$  becomes a bootstrap equation for the double-copy kernel.\footnote{In  principle, a kernel with non-minimal rank is allowed by the KLT bootstrap method, but there is evidence that spurious poles arise in general $d$ dimensions \cite{Johnson:2020pny,Chi:2021mio}; for $d=3$, see \cite{Gonzalez:2021bes}.} 
We now summarize the results of  Ref.~\cite{Chi:2021mio} for 4- and 5-point. 

\vspace{1mm}
\noindent {\bf 4-point.} 
At 4-point, all $3! \times 3!$ entries of $\mathbf{m}_n$ can be expressed in terms of just three functions\footnote{We define  
$s_{ij}=(p_i+p_j)^2$ and $s=s_{12}$, $t=s_{13}$, $u=s_{14}$. At 4-point,  $s+t+u=0$.}  
\bea
  \nonumber
  \hspace{-2mm}
  &&f_1(s,t) = m_4[1234|1234]\,,  ~~ f_2(s,t) = m_4[1234|1243]\,,
  \\  
  \hspace{-2mm}
  &&f_6(s,t) = m_4[1234|1432]\,
\eea
via cyclicity and momentum relabeling, and the minimal rank 1 condition is solved by 
\be
  \label{bootstrapf2}
  \begin{split}
  &f_6(s,t) = f_1(s, t) = \frac{f_2(s, t) f_2(u, s)}{f_2(t, s)}\,,\\
  &f_2(s, t) f_2(u, s) f_2(t,u) = f_2(t, s)  f_2(u, t) f_2(s, u)\,.
  \end{split}
\ee
These equations ensure that $f_1$ is cyclic, $f_1(u, t) = f_1(s, t)$. 
These conditions are solved by the tree amplitudes of BAS in Eq.~\reef{BAS} 
and by the inverse string kernel of Ref.~\cite{Mizera:2016jhj}, respectively
\be
\label{BASstr}
 f_2^\text{BAS} = -\frac{1}{s}\,,~~~~
 f_2^\text{str}(s,t) = -\frac{\pi \alpha'}{\sin(\pi \alpha' s)}\,.
\ee

Since on-shell local operators are in one-to-one correspondence with local on-shell matrix elements, the most general ansatz for the BAS EFT is given by higher-order polynomial terms in $f_2$ as
\be 
  \label{f2ansatz}
   f_2(s,t) = -\frac{1}{s} 
   + \sum_{k=0}^N \sum_{r=0}^k
   a_{k,r} \,s^r \,t^{k-r} \, .
\ee
Here all possible local operators up to $2N$ derivatives are included. The Wilson coefficients $a_{2i,2i}$ must be set to zero for all $i=0,1,2,\ldots$ to avoid unphysical poles in $f_1$ via Eq.~\reef{bootstrapf2}.
The 4-point bootstrap equation, Eq.~\reef{bootstrapf2}, is solved order by order in the Mandelstam variables and for the first few orders we find 
\be\nonumber
\label{4ptres}
  \begin{array}{cll}
  k&  \text{constraints}\\
  \hline
  1  & \text{none} \\[0.1mm]
  2  & a_{2, 1} = a_{2, 0} \\[0.1mm]
  3  &  \text{none} \\[0.1mm]
  4  &  a_{4, 3} = a_{4, 0} - a_{4, 1} + a_{4, 2}  \\[0.1mm]
  5  & a_{5, 4} = a_{5, 0} - a_{5, 1} + a_{5, 3}
  + 
 a_{1, 0} a_{1, 1} \big(a_{1, 0} - a_{1, 1} \big)\\[0.1mm]
  &~~~+ a_{1, 1} (a_{3, 1} - a_{3, 2}\big) - 
 a_{1, 0}\big(a_{3, 0} - a_{3, 2} + a_{3, 3}\big).\\[0.1mm]
% 6 & a_{6, 5} = a_{6, 0} - a_{6, 1} + a_{6, 2} - a_{6, 3} + a_{6, 4}
  \end{array}
\ee 
Via Eq.~\reef{mandSinv}, the amplitudes of this  BAS EFT define a kernel that is much more general than the string kernel. To match the inverse string kernel $f_2^\text{str}$ in Eq.~\reef{BASstr}, set all $a_{k,r} = 0$ except for 
$a_{k,k}$ with $k$ odd for which 
\be
 \label{astring}
 \begin{split}
  &a_{1,1} = -\tfrac{1}{6}\pi^{2} \alpha'^{2}\,,\\
  ~~
  &a_{3,3} = -\tfrac{7}{360}\pi^4 \alpha'^4\,
  ~~
  a_{5,5} =-\tfrac{31}{15120}\pi^6 \alpha'^6\,,
  \ldots\ ~~~
  \end{split}
\ee
From a bottom-up perspective, we need to fix an infinite number of coefficients to non-zero values.

\begin{table}[t]
\be\nonumber
 \begin{array}{|r|rrr|l|}
 \hline
 k  & \text{4-pt}
   & \text{5-pt}
   & \text{6-pt} 
   ~&~ \text{free after}
   \\
   & O(s^{k})
   & O(s^{k-1})
   & O(s^{k-2}) 
   ~&~\text{6-pt boostrap}\\
   \hline
   1 & 2 & 0 & 0 ~&~ 2 ~~a_{1,0},a_{1,1}\\
   2 & 1 & 0 & 24  ~&~ 1  ~~a_{2,0}\\
  3 & 4 & 0 & 216 ~&~ 0 \\
  4 & 3 & 4 & 1080 ~&~ 1 ~~a_{4,0} \\
  5 & 5 & 10 & 3960 ~&~ 0 \\
 \hline
 \end{array}
\ee
\caption{\label{tab:count}Number of free parameters at each order in Mandelstams in the BAS + EFT ansatz before and after the 6-point bootstrap. 
The counting in the table refers to parameters left free after the 4-point and 5-point minimal rank bootstrap. 
For example,  for $k=4$, the 3 free parameters at 4-point are $a_{4,0},a_{4,1},a_{4,2}$, because $a_{4,3}$ was fixed by the 4-point bootstrap and we needed $a_{4,4} = 0$ to ensure locality of $f_1$.  
The 6-point bootstrap condition fixes all 3+4+1080 parameters except 1, namely $a_{4,0}$.}
\end{table}

\vspace{1mm}
\noindent {\bf 5-point.} 
The 5-point KLT bootstrap 
requires $\mathbf{m}_5$ to have rank 2, which significantly
constrains the Wilson coefficients of the local 5-point interactions. Since the BAS EFT amplitudes $m_5$ depend on $m_4$ through factorization, it is noteworthy that no additional constraints (as explicitly checked up to $O(s^8)$) arise on the 4-point Wilson coefficients $a_{k,r}$.

\section{New Results at 6-Point}
\label{s:6pt}

New constraints on the 4-point double-copy kernel arise from 6-point KLT bootstrap. The 6-point analysis is done by first using cyclicity and momentum relabeling to parameterize the $(6-1)! = 120$ distinct $m_6[123456|b]$ amplitudes in terms of 24 ``basis" amplitudes. All pole-terms of these 24 basis amplitudes are then fixed by their factorization to the known 4- and 5-point amplitudes $m_4$ and $m_5$ of the BAS EFT. Local 6-point terms are included as all possible polynomial terms in a choice of the 9 Mandelstam variables that are independent under 6-point momentum conservation. Table \ref{tab:count} gives the parameter count at each order.

We have imposed minimal rank 6 on the $120 \times 120$ matrix of doubly-color ordered 6-point amplitudes of BAS EFT by setting the $7 \times 7$ minors to zero up to and including cubic order $\mathcal{O}({s^3})$.
This constrains the 4-point coefficients $a_{k,r}$ up to and including $\mathcal{O}(s^{5})$ and the full results are summarized in Table \ref{tab:results}. In Section \ref{s:hybrid}, we extend these results to higher order. 
The result is that at 4-point, all parameters are fixed except 
\be
  \label{4ptparameters}
  a_{1,0}\,, a_{1,1}~~\text{and}~~ a_{2k,0}~\text{for}~
  k=1,2,3,\ldots 
\ee
for which we are free to choose any values.

\begin{table}[t!]
\be\nonumber
  \begin{array}{|cll|}
  \hline
a_{1,0} && \text{free}\\
a_{1,1} && \text{free}\\[2mm]
a_{2,0} && \text{free} \\
a_{2,1} && a_{2,0} \\[2mm]
a_{3, 0}&=& \frac{2}{5 } a_{1, 0} \big(a_{1, 0} - 2 a_{1, 1}\big) \\[1mm]
a_{3, 1} &=& \frac{1}{10 } a_{1, 0} \big(a_{1, 0} - 12 a_{1, 1}\big) \\[1mm]
a_{3, 2} &=& \frac{1}{5 } a_{1, 0} \big(2 a_{1, 0} - 9 a_{1, 1}\big)\\[1mm]
a_{3, 3} &=& -\frac{7}{10 } a_{1, 1}^2\\[2mm]
a_{4,0} && \text{free}\\[1mm]
a_{4, 1} &=& - a_{1, 0}\, a_{2, 0}  + 2 a_{4, 0} \\[1mm]
a_{4, 2} &=& -\big(a_{1, 0} + a_{1, 1}\big)a_{2, 0} + 2  a_{4, 0}\\[1mm]
a_{4, 3} &=& - a_{1, 1} \, a_{2, 0}  +  a_{4, 0}\\[2mm]
a_{5,0} &=&
\frac{8}{35 }
 a_{1,0} \big(a_{1, 0}^2 - 3 a_{1,0}\, a_{1, 1} + 3 a_{1,1}^2\big)\\[1mm]
 a_{5,1} &=&
\frac{2}{35 }
 a_{1,0} \big(3a_{1, 0}^2 - 16 a_{1,0}\, a_{1, 1} + 30 a_{1,1}^2\big) - \frac{a_{2,0}^{2}}{2} \\[1mm]
 a_{5,2} &=&
\frac{1}{70 }
 a_{1,0} \big(23a_{1, 0}^2 - 104 a_{1,0}\, a_{1, 1} + 216 a_{1,1}^2\big) - a_{2,0}^{2}\\[1mm]
  a_{5,3} &=&
\frac{1}{70 }
 a_{1,0} \big(12a_{1, 0}^2 - 71 a_{1,0}\, a_{1, 1} + 204 a_{1,1}^2\big) - \frac{a_{2,0}^{2}}{2}\\[1mm]
 a_{5,4} &=&
\frac{1}{70 }
 a_{1,0} \big(16a_{1, 0}^2 - 76 a_{1,0}\, a_{1, 1} + 153 a_{1,1}^2\big) \\[1mm]
  a_{5,5} &=&\frac{31}{70 } a_{1, 1}^3\\[1mm]
  \hline
\end{array}
\ee
\caption{\label{tab:results}Combined results of the KLT bootstrap at 4-, 5-, and 6-point for the coefficients up to $\mathcal{O}(s^5)$ in  the 4-point amplitude $f_2$ in Eq.~\reef{f2ansatz}.}
\end{table}
The generalized KKBCJ relations arising from Eq.~\reef{KLTalgebra0cond} can all be written in terms of ratios of $f_2$. For example one of the relations required by $\text{L}\otimes \mathbb{1} = \text{L}$ is 
\be
\label{lgenkkbjrel}
 A^\text{L}_{4}[1234] =\frac{f_2(u,s)}{f_2(t,s)}A^\text{L}_{4}[1243] \,.
\ee 
The ratios of $f_2$ turn out to be independent of all $a_{2k,0}$. In fact, the generalized L-sector KKBCJ relations only depend on a single parameter, namely $a_{1,1} - a_{1,0}$ whereas the relations from $\mathbb{1} \otimes  \text{R} =  \text{R}$ only depend on $a_{1,1}$. Writing 
\be
\label{aLaR}
a_{1,1}=-\frac{\pi^{2}\alpha_\text{R}'^{2}}{6}, \quad a_{1,0}=\frac{\pi^{2}}{6}(\alpha_\text{L}'^{2}-\alpha_\text{R}'^{2})\,,
\ee
 our results for the L and R generalized KKBCJ relations can be identified precisely as the low-energy expansion of the string monodromy relations with separate L or R choices of $\alpha'$. Depending on whether $\alpha'_\text{L,R}$ are zero or not means that that generalized 4-point KKBCJ relations are then either the field theory KKBCJ relations or the string monodromy relations!
 
 We now connect the results of the KLT bootstrap to known special cases.

\vspace{1mm}
\noindent {\bf Inverse String Kernel.} The inverse string kernel has monodromy relations with the same $\alpha'$ for the two color-orderings, i.e.~$\alpha'_\text{L}=\alpha'_\text{R}$. By Eq.~\reef{aLaR}, this choice requires $a_{1,0}= 0$. Table \ref{tab:results} then shows that almost all coefficients vanish except the $a_{2k,0}$'s and   $a_{k,k}$ for $k$ odd, which directly give the string kernel coefficients in Eq.~\reef{astring}. Thus, 
\be
 \label{strchoice}
 \alpha' \equiv \alpha'_\text{L}=\alpha'_\text{R}\, ~~\text{and}~~
 a_{2k,0}=0\,
\ee
matches the 
inverse string kernel. 

\vspace{2mm}
\noindent {\bf Hybrid Models GF (Generalized Z-theory).}  
Ref.~\cite{Chen:2022shl} studied BAS EFTs with tree amplitudes that satisfy the field theory KKBCJ relations on the second color-ordering. These relations are imposed as $(n-1)!-(n-3)!$ null vector conditions on the matrix $\mathbf{m}_n$. Hence, it has rank $(n-3)!$ and  must be in the same class as the inverse double-copy kernels studied in this paper. We call these models {\em hybrid models} and denote them by {\bf GF} to indicate that they have {\bf G}eneral (i.e.~no imposed constraints) on the first color-structure and obey {\bf F}ield theory 
KKBCJ relations on the second color-structure. It was shown in Ref.~\cite{Chen:2022shl} that at 4-point the generalized KKBCJ relations of the first color-structure (i.e.~G) are the string monodromy relations with $a_{1,0}$ simply a choice of the scale of $\alpha'$.

We obtain the GF models of Ref.~\cite{Chen:2022shl} by setting 
\be
  a_{1,1} = a_{1,0}\,,
  ~~~~
  \text{i.e.}
  ~~
  \alpha'_\text{L} = 0\,.
\ee
Additionally choosing 
\be
  \label{ZtheoryCoeff}
 a_{2k,0} = - \zeta(2k+1) \,\alpha'^{2k+1}\,,
 \ee
 where $\alpha' = \alpha'_\text{R}$ and $\zeta(p)$ is the Riemann Zeta function, we match the 4-point amplitudes of {\bf Z-theory} in Ref.~\cite{Broedel:2013tta}. 
It is useful to note that the open string tree amplitudes can be obtained as the double copy
\be
  \label{openisZxYM}
  \text{open string tree} = \text{Z} \otimes_\text{FF} \text{YM} \,,
\ee
where $\otimes_\text{FF}$ indicates the field theory kernel with no higher-derivative corrections.

\vspace{2mm}
\noindent {\bf Closed String $J$-integrals.} 
The closed string tree amplitudes can be written in terms of period integrals, the ``J-integrals", as a field theory kernel double copy \cite{Stieberger:2014hba,Mizera:2017cqs,Schlotterer:2018zce,Vanhove:2018elu,Brown:2018omk}
\be
  \label{Jint}
  \text{closed string tree} = \text{YM}
  \otimes_\text{FF}
  \text{J}
  \otimes_\text{FF}
  \text{YM}\,.
\ee
The J-integral amplitudes are a special case of the BAS EFT amplitudes, namely those that obey the field theory KKBCJ relations on both color-structures, i.e.~$\alpha'_\text{L}=\alpha'_\text{R} = 0$. 
We obtain the J-integral amplitudes by additionally choosing  $a_{2k,0} = - 2\zeta(2k+1) \alpha'^{2k+1}$.

\section{Hybrid Decomposition}
\label{s:hybrid}
The process of solving the vanishing conditions for $7 \times 7$ minors of $\mathbf{m}_6$ becomes increasingly difficult at higher orders in the derivative expansion. 
However, using the {\em  hybrid decomposition conjecture} of Ref.~\cite{ACHElater} we are able to go to much higher order. This conjecture posits that 
the most general BAS EFT (denoted GG) whose tree amplitude matrix $\mathbf{m}_n$ has minimal rank $(n-3)!$ can be obtained by a field theory double-copy of two hybrid models: 
\be
  \label{hybridconj}
  \text{GG} = \text{GF}
  \otimes_\text{FF} \text{FG}\,.
\ee
Here GF is the hybrid model described above and FG is the hybrid model with first and second color orders interchanged. GF and FG have independent coefficients, $a^\text{GF}_{k,r}$ and $a^\text{FG}_{k,r}$. 
The subscript FF on the product in Eq.~\reef{hybridconj} emphasizes that the double copy is done with the standard field theory kernel. 

At 4-, 5-, and 6-points, one can directly test Eq.~\reef{hybridconj} using the results for GG presented in Section~\ref{s:6pt} and those for GF in Ref.~\cite{Chen:2022shl}. Further evidence will be presented in 
Ref.~\cite{ACHElater}. 
The key point here is that it is much easier to solve the linear field theory KKBCJ constraints for GF than it is to impose vanishing $7\times 7$ determinant conditions for GG. 
The hybrid model GF was solved to order 18 in the Mandelstam expansion at 4-point with constraints from 6-point KKBCJ relations in Ref.~\cite{Chen:2022shl},  and a closed-form expression for $f^\text{GF}_2$ was proposed; see in Eq.~(27) of Ref.~\cite{Chen:2022shl}.

Based on the closed-form expression for $f^\text{GF}_2$  and the hybrid conjecture Eq.~\reef{hybridconj}, we propose the following re-summed result for $f^\text{GG}_{2}$:
\be \label{finalsolu}
   f^\text{GG}_2(s,t) =  f_2^{(0)}(s,t)\,U(s,t,u) \,,
\ee
where 
{\small
\bea
   \nonumber
    \hspace{-2mm}
    &&f_{2}^{(0)}=-\frac{\pi}{s}\sqrt{\frac{\alpha'_\text{R}\alpha'_\text{L}\,s^{2}\sin(\pi\alpha_\text{R}'t)
    \sin(\pi\alpha_\text{L}'u)}{\sin(\pi\alpha_\text{R}'s)\sin(\pi\alpha_\text{R}'u)\sin(\pi\alpha_\text{L}'s)
    \sin(\pi\alpha_\text{L}'t)
    }}\ , 
\\
    \hspace{-2mm}&&
    \log U
    = \sum_{k=1}^\infty \frac{a_{2k,0}}{2k+1}  \Big(s^{2k+1}+t^{2k+1}+u^{2k+1}\Big) \ .
\label{exponform}
\eea}%
Here, $a_{2k,0} = a^\text{GF}_{2k,0}+a^\text{FG}_{2k,0}$, and $\alpha_\text{R/L}'$ are the $\alpha'$'s associated with the 4-point string monodromy relations of the GF and FG models, respectively. Using the definitions in Eq.~\reef{aLaR}, the low-energy expansion of Eqs.~\reef{finalsolu}-\reef{exponform} matches  the results in Table \ref{tab:results}. When $\alpha_\text{R}'=\alpha_\text{L}'$ and $a_{2k,0}=0$ we recover the inverse string kernel $f^\text{str}$ in Eq.~\reef{BASstr} from Eqs. \reef{finalsolu}-\reef{exponform}.

The closed form expression in Eq.~\reef{exponform} has an infinite tower of unphysical poles, so it should be understood in the low-energy expansion. 

Additional constraints on the 4-point coefficients $a_{k,r}$ from a higher-point bootstrap are unlikely. Conditions on $a_{2k,0}$ are expected to be impossible because Z-theory has $a_{2k,0}$ fixed to $\zeta(2k+1)$ and there are (conjecturally) no polynomial conditions that can relate these odd-argument  transcendental numbers \cite{2000math.....10140H,Schlotterer:2012ny,Broedel:2013tta,Brown:2013gia}. It is consistent with the known cases in Sec.~\ref{s:6pt} that the  higher-point bootstrap could give a condition like $\alpha_\text{L}'\alpha_\text{R}'(\alpha_\text{L}'-\alpha_\text{R}') =0$, however, this would be incompatible with the proposed hybrid decomposition.

%%%%%%%%%%%%%%%%
\section{Implications for KLT Double Copy}
\label{s:impli}

\noindent {\bf Monodromy Relations.}
The closed-form expression in Eqs.~\reef{finalsolu}-\reef{exponform} explains why the dependence on $a_{2k,0}$ disappears from the 4-point generalized KKBCJ relations: they only depend on ratios of $f_2$, such as in Eq.~\reef{lgenkkbjrel}, hence the symmetric function $U$ cancels from these expressions. Moreover, in the hybrid decomposition construction, the monodromy relations from the hybrid models are inherited by GG in \reef{hybridconj}. For example, we directly see how Eq.~\reef{lgenkkbjrel} gives the string monodromy relation
\begin{equation}
\sin(\pi \alpha_{L}'u)
\,A^\text{L}_{4}[1234]=\sin(\pi \alpha_{L}'t)\,
A^\text{L}_{4}[1243]\ .
\end{equation}
Thus, both the low-energy expansion in Sec.~\ref{s:6pt} and the construction via the hybrid decomposition in Sec.~\ref{s:hybrid} offer evidence that the generalized KKBCJ relations of the minimal double-copy kernel at 4-point must be either the string monodromy relations ($\alpha_\text{L/R}$ non-zero) or the field theory KKBCJ relations ($\alpha_\text{L/R}$  zero). It is interesting that such stringy properties arise from the basic assumptions of the KLT algebra!

\vspace{1mm}
\noindent {\bf Kernel Equivalence.} 
Even though $U$ in Eq.~\reef{finalsolu} does not enter the generalized KKBCJ conditions, such as in Eq.~\reef{lgenkkbjrel}, it does contribute to the final double result. However, any contribution from the $U$ in the kernel can be absorbed into the input amplitudes. Concretely, if we use $U$ to redefine the Wilson coefficients of the higher-derivative terms of the L or R sector amplitudes, e.g.~${A}_4^{\textrm{R}}[b] ~\to~  U\, {A}_4^{\textrm{R}}[b]$ for all color orders $b$ and simultaneously rescale $m_4[a|b] \to m_4[a|b] U$, the result of the double-copy is unchanged:
\begin{equation}
\label{KLTwU}
  \begin{split}
    &{A}_4^{\text{L}\otimes \text{R}} 
    =    {A}_4^{\textrm{L}}[a]\,
    S_4[a|b] \,
    {A}_4^{\textrm{R}}[b] \\
    &\quad \to 
        {A}_4^{\textrm{L}}[a]\,
    \big(S_4[a|b] U^{-1}\big) \,
    \big(U {A}_4^{\textrm{R}}[b]\big)
    = {A}_4^{\text{L}\otimes \text{R}} 
    \,.
  \end{split}
\end{equation}
This means that $f_2$ and $f_2 U$ are functionally equivalent from the low energy perspective. These statements are valid in the EFT context, but not necessarity for  closed expressions because a resummed version of $U$ could introduce unwanted spurious poles. 

\vspace{1mm}
\noindent {\bf Single-Valued Projection.}
With the closed string tree amplitudes being the double copy of the open string using the string kernel, denoted $\otimes_{\alpha'\!\alpha'}$, we have from Eq.~\reef{openisZxYM} the identity
\be
  \text{closed string tree}
  = \text{YM} 
  \otimes_\text{FF} 
  \text{Z}^T 
  \otimes_{\alpha'\!\alpha'} 
  \text{Z}
  \otimes_\text{FF} 
  \text{YM}\,,
\ee
where Z$^T$ is the model whose tree amplitudes are $(\mathbf{m}_n^\text{Z})^T$. 
Comparing with Eq.~\reef{Jint} gives 
\cite{Stieberger:2014hba,Mizera:2017cqs,Schlotterer:2018zce,Vanhove:2018elu,Brown:2018omk}
\be
   \text{J} = 
   \text{Z}^T 
  \otimes_{\alpha'\!\alpha'} 
  \text{Z}\,.
\ee
Now, Eq.~\reef{finalsolu} is valid for the string kernel, with $U=1$ and for Z-theory whose $U_\text{Z}$ is given by \reef{exponform} with coefficients \reef{ZtheoryCoeff}. At 4-point, the $J$-integrals have $f_2^{(0)} = -1/s$, so all the $\alpha'$-dependence from the $f_2^{(0)}$'s of Z-theory and the string kernel must cancel: this eliminates  $\pi$. Only $U_\text{Z}^2$ remains, which has parameters $2 a^\text{Z}_{2k,0}$ due to its exponentiated form in Eq.~\reef{exponform}. 
Thus at 4-point, Eqs.~\reef{finalsolu}-\reef{exponform} reproduce the ``single-valued projection" (sv) \cite{Schnetz:2013hqa,Brown:2013gia}:
\be 
  \begin{split}
  &\text{J} = \text{sv(Z)} \,,\\
  ~~~
  &\text{sv:}~~\zeta(\text{even}) \to 0\,,
  ~~~\zeta(\text{odd}) \to 2\zeta(\text{odd})\,.
  \end{split}
\ee
This implies that  
the closed (super)string tree amplitude can be obtained as 
sv(open)$\otimes_\text{FF}$(S)YM \cite{Schlotterer:2012ny,Stieberger:2013wea}.

The generalization  to  $\alpha'_\text{L} \ne \alpha'_\text{R}$ is straightforward. 
It would be interesting to understand if the double-copy bootstrap can similarly provide an explanation of the single-valued projection at higher point. 

\vspace{1mm}
\noindent {\bf KLT vs.~BCJ.}
An alternative representation of the double-copy kernel with higher-derivative corrections similar to the BCJ representation of Ref.~\cite{Bern:2008qj} was proposed in Refs.~\cite{Carrasco:2019yyn,Carrasco:2021ptp}. Ref.~\cite{Bonnefoy:2021qgu} studied the connection between these constructions and found that the KLT bootstrap is equivalent to the higher-derivative BCJ construction at 4-point and checked the equivalence of the constructions at 5-point up to $\mathcal{O}(s^{3})$. However, the higher-derivative BCJ representation in Ref.~\cite{Carrasco:2019yyn,Carrasco:2021ptp} can be truncated at finite order in Mandelstams. The results in this Letter suggest that the higher-derivative BCJ representation in Refs.~\cite{Carrasco:2019yyn,Carrasco:2021ptp} are in tension with locality at 6-point and that an infinite number of 4-point higher derivative operators are necessary to resolve this tension. Results along these lines in the context of YM have been discussed in Refs.~\cite{Carrasco:2022lbm,Carrasco:2022sck}.

\vspace{2mm}
We assumed the BAS EFT to have a non-vanishing cubic interaction term. This  ensured that the usual field theory double-copy was obtained from the low-energy limit of the generalized double-copy. Relaxing this condition may result in a different form of the double-copy.

\medskip

\section{Acknowledgements.}
We would like to thank Nima Arkani-Hamed, Justin Berman, Lance Dixon, Caroline Figueiredo, Nick Geiser, Alfredo Guevara Gonzalez,  Aaron Hillman, Callum Jones, Sebastian Mizera, Shruti Paranjape, and Fei Teng for useful comments and discussions. HE and ASC are supported in part by DE-SC0007859. ASC is also supported in part by a Leinweber Summer Fellowship. AH was supported by a Rackham Predoctoral Fellowship from the University of Michigan.

\bibliography{apssamp}
\end{document}